\documentclass[12pt,a4paper]{article}
\usepackage{a4wide}
\usepackage{amsfonts}
\begin{document}
{\renewcommand{\thefootnote}{\fnsymbol{footnote}}
\hfill  PITHA -- 00/17\\
\medskip
\hfill gr--qc/0008052\\
\medskip
\begin{center}
{\LARGE  Loop Quantum Cosmology III:\\\smallskip Wheeler--DeWitt Operators}\\
\vspace{1.5em}
Martin Bojowald\footnote{e-mail address:
{\tt bojowald@physik.rwth-aachen.de}}\\
Institute for Theoretical Physics, RWTH Aachen\\
D--52056 Aachen, Germany\\
\vspace{1.5em}
\end{center}
}

\setcounter{footnote}{0}

\newtheorem{lemma}{Lemma}

\newcommand{\proofend}{\raisebox{1.3mm}{\fbox{\begin{minipage}[b][0cm][b]{0cm}
\end{minipage}}}}
\newenvironment{proof}{\noindent{\it Proof:} }{\mbox{}\hfill 
  \proofend\\\mbox{}}

\newcommand{\AbB}{\overline{{\cal A}}_B}
\newcommand{\UbB}{\overline{{\cal U}}_B}
\newcommand{\AbS}{\overline{{\cal A}}_{\Sigma}}
\newcommand{\Haux}{{\cal H}_{{\mathrm{aux}}}}
\newcommand{\Hdiff}{{\cal H}_{{\mathrm{diff}}}}
\newcommand{\HE}{{\cal H}^{({\mathrm E})}}
\newcommand{\hatHE}{\hat{{\cal H}}^{({\mathrm E})}}
\newcommand{\SF}{\overline{S/F}}
\newcommand{\SFe}{\overline{S/F}\mbox{}_{\epsilon}}
\newcommand{\lP}{l_{{\mathrm P}}}

\newcommand{\ts}{\textstyle}
\newcommand{\half}{{\textstyle\frac{1}{2}}}
\newcommand{\md}{{\mathrm{d}}}
\newcommand{\Aut}{\mathop{\mathrm{Aut}}}
\newcommand{\Ad}{\mathop{\mathrm{Ad}}\nolimits}
\newcommand{\ad}{\mathop{\mathrm{ad}}\nolimits}
\newcommand{\Hom}{\mathop{\mathrm{Hom}}}
\newcommand{\Ima}{\mathop{\mathrm{Im}}}
\newcommand{\id}{\mathop{\mathrm{id}}}
\newcommand{\diag}{\mathop{\mathrm{diag}}}
\newcommand{\Kern}{\mathop{\mathrm{ker}}}
\newcommand{\tr}{\mathop{\mathrm{tr}}}
\newcommand{\sgn}{\mathop{\mathrm{sgn}}}
\newcommand{\semidir}{\mathrel{\mathrm{\times\mkern-3.3mu\protect%
\rule[0.04ex]{0.04em}{1.05ex}\mkern3.3mu\mbox{}}}}
\newcommand{\dive}{\mathop{\mathrm{div}}}
\newcommand{\Diff}{\mathop{\mathrm{Diff}}\nolimits}

\newcommand*{\R}{{\mathbb R}}
\newcommand*{\N}{{\mathbb N}}
\newcommand*{\Z}{{\mathbb Z}}
\newcommand*{\Q}{{\mathbb Q}}
\newcommand*{\C}{{\mathbb C}}

\begin{abstract}
  In the framework of loop quantum cosmology anomaly free
  quantizations of the Hamiltonian constraint for Bianchi class A,
  locally rotationally symmetric and isotropic models are given. Basic
  ideas of the construction in (non-symmetric) loop quantum gravity
  can be used, but there are also further inputs because the special
  structure of symmetric models has to be respected by operators. In
  particular, the basic building blocks of the homogeneous models are
  point holonomies rather than holonomies necessitating a new
  regularization procedure. In this respect, our construction is
  applicable also for other (non-homogeneous) symmetric models, e.g.\ 
  the spherically symmetric one.
\end{abstract}

\section{Introduction}

Loop quantum gravity \cite{Rov:Loops}, a program for the canonical
quantization of general relativity, has lead to a well-understood
theory of quantum geometry \cite{AshKras} predicting discrete spectra
for geometrical operators like area or volume
\cite{AreaVol,Loll:Vol,Area,Vol2,Vol,RecTh}. The main open problems
concern the semiclassical limit and the dynamics of the quantum
theory. Recently, a strategy to attack the first problem has been
proposed \cite{GCSI} by constructing coherent states which are peaked
around a given classical configuration.

As for dynamics, there are a number of anomaly free quantizations of the
Hamiltonian constraint \cite{QSDI,Pullin} and a class of very special
solutions \cite{QSDIII}. It is, however, not clear how to interpret
any possible solution, in part because of conceptual problems, most
importantly the problem of time, which are already present in the
classical canonical formulation. These classical conceptual questions
have been mainly investigated in reduced models subject to a symmetry
condition such that one has a good control over all solutions. Most
prominent in this context are cosmological models which are
homogeneous in space and present examples of mini-superspaces which
after reduction can be quantized similar to a conventional mechanical
system \cite{DeWitt,Misner}. In the first part of this series
\cite{cosmoI} a way was proposed how to perform such a symmetry
reduction at the kinematical level of loop quantum gravity based on
the general concept of symmetric states in diffeomorphism invariant
quantum theories of connections \cite{SymmRed}. Equivalently, this can
be seen as a loop quantization of a mini-superspace, but the quantum
states can always be interpreted as generalized, symmetric states of
loop quantum gravity. Based on the kinematical level, the next step is
to quantize the reduced Hamiltonian constraint of a given model
resulting in the Wheeler--DeWitt operator which governs its
dynamics. Having such a quantization one can then look whether the
classical conceptions regarding the problem of time apply and what
they imply for the quantum dynamics \cite{PhD,cosmoIV}.

Quantizations of the Hamiltonian constraints for some models are the
subject of the present paper. Because the aim is to understand the
dynamics of loop quantum gravity, we will follow the basic
construction in Ref.\ \cite{QSDI} as close as possible. Although the
ensuing Wheeler--DeWitt operators will be quite similar to the one of
the full theory, there is additional input needed in their
quantization, essentially due to the following basic ingredient: in
order to obtain the curvature components present in the classical
constraints, one has to lay a loop in the space manifold whose
holonomy approximates the curvature of the Ashtekar connection. In the
full theory, one then couples the classical limit with a continuum
limit in which all these infinitesimal loops exactly reproduce the
curvature components. Moreover, one has to use diffeomorphism
invariance in order to ensure that the continuum limit does not affect
the quantum constraint. In homogeneous models we then have to face the
following problems: First, there is no place for such loops because
the reduced models are formulated in a single point and their quantum
states are based on point holonomies rather than ordinary holonomies.
Therefore, a new regularization is needed which also provides a
substitute for the continuum limit which played an important
role in the quantization of the Hamiltonian constraint in the full
theory. We will do this by extending the point, in which the reduced
model is formulated, to an auxiliary manifold using the structure of
the symmetry group. This step has already been done in the
construction of derivative operators \cite{cosmoI}. The next problem
then is to lay the loops in such a way that the Wheeler--DeWitt
operator respects the symmetry, resulting in the main difference to
the operator in the full theory \cite{QSDI}.

We will start by reviewing the structure of loop quantum cosmological
models in Section \ref{s:Models} and the construction of the
Wheeler--DeWitt operator in the full theory in Section
\ref{s:Full}. Then we will be ready to quantize the Hamiltonian
constraint of Bianchi class A models (Section \ref{s:Bianchi}) and
subsequently of locally rotationally symmetric and isotropic models
(Section \ref{s:Iso}).

\section{Bianchi class A, LRS and Isotropic Models}
\label{s:Models}

In this section we recall the results of a symmetry reduction for
cosmological models and their quantization at the kinematical level.
For details we refer to Ref.\ \cite{cosmoI}.

Bianchi models are homogeneous models which have a symmetry group $S$
with structure constants $c^K_{IJ}$ acting freely and transitively on
the space manifold $\Sigma$. Bianchi class A models are characterized
by $c^I_{JI}=0$. Some of them can be reduced by requiring a
non-trivial isotropy group which is isomorphic either to $U(1)$
(locally rotationally symmetric, LRS models) or to $SO(3)$ (isotropic
models). In all cases, $S$ is a semidirect product of a translation
group $N$ and an isotropy group $F$. We are interested in the
reduction of general relativity formulated in real Ashtekar variables
and, therefore, need the general expressions of invariant
$SU(2)$-connections and ${\cal L}SU(2)$-valued density weighted
dreibein fields.

An invariant connection can always be decomposed as
$A=\phi^i_I\omega^I\tau_i$ using the generators
$\tau_j=-\frac{i}{2}\sigma_j$ (with Pauli matrices $\sigma_j$) of
$G=SU(2)$ and left invariant one-forms $\omega^I$ on the translation
group $N$ (which is identical to $S$ for Bianchi models). Invariant
connections are parameterized by the components $\phi^i_I$ of a linear
map $\phi\colon{\cal L}N\to{\cal L}G$ which is arbitrary for general
Bianchi models, but restricted to be of the form
\begin{equation} \label{scalar}
 \phi^i_1 = 2^{-\frac{1}{2}}(a\Lambda^i_1+b\Lambda^i_2)\quad,\quad
 \phi^i_2 = 2^{-\frac{1}{2}}(-b\Lambda^i_1+a\Lambda^i_2)\quad,\quad
 \phi^i_3 = c\Lambda^i_3
\end{equation}
for LRS models and $\phi_I^i=c\Lambda^i_I$ for isotropic
models. $\Lambda^i_I$ is a dreibein which is rotated by the internal
$SU(2)$-gauge transformations.

Similarly, an invariant density weighted dreibein field can be
decomposed as $E^a_i=\sqrt{g_0}\,p^I_iX^a_I$ with left invariant vector
fields on $N$ satisfying $\omega^I(X_J)=\delta^I_J$ and the
determinant $g_0$ of a left invariant metric on $N$ defined by
$\omega^1\wedge \omega^2\wedge \omega^3= \sqrt{g_0}\,\md^3x$. Again, for
Bianchi models the coefficients $p^I_i$ are arbitrary, but restricted
to be of the form
\[
  p^1_i=2^{-\frac{1}{2}}(p_a\Lambda^i_1+p_b\Lambda^i_2)\quad,\quad
  p^2_i=2^{-\frac{1}{2}}(-p_b\Lambda^i_1+p_a\Lambda^i_2)\quad,\quad 
  p^3_i=p_c\Lambda^i_3
\]
for LRS models and $p^I_i=p\Lambda^I_i$ for isotropic models.

The symplectic structure is given by
\[
  \{\phi^i_I,p^J_j\}=\kappa\iota'\delta^i_j\delta^J_I
\]
for Bianchi models,
\[
  \{a,p_a\}=\{b,p_b\}=\{c,p_c\}=\kappa\iota'
\]
and vanishing in all other cases for LRS models, and
$\{c,p\}=\kappa\iota'$ for isotropic models. Here,
$\kappa=8\pi G$ is the gravitational constant and $\iota':=\iota V_0^{-1}$
the reduced Immirzi parameter with
$V_0:=\int_{\Sigma}\md^3x\sqrt{g_0}$.

Reduced expressions can be derived by inserting invariant connections
and dreibein fields into the unreduced expressions of the full theory.
In this paper we need the Euclidean parts of the reduced Hamiltonian
constraints, which are
\begin{eqnarray}
  g_0^{-1}{\cal H}^{({\mathrm E})} & = & g_0^{-1}\epsilon_{ijk}
  F^i_{IJ}E^I_jE^J_k=
  -\epsilon_{ijk} c^K_{IJ}\phi^i_Kp^I_jp^J_k+
  \epsilon_{ijk}\epsilon_{ilm} \phi^l_I\phi^m_Jp^I_jp^J_k\nonumber\\
  & = & -\epsilon_{ijk}c^K_{IJ} \phi^i_Kp^I_jp^J_k+
  \phi^j_I\phi^k_Jp^I_jp^J_k-
  \phi^k_I\phi^j_Jp^I_jp^J_k   \label{hambianchi}
\end{eqnarray}
for Bianchi models,
\begin{eqnarray}
  g_0^{-1}{\cal H}^{({\mathrm E})}& = & -(n^{(1)}+n^{(2)})(ap_a+bp_b)p_c-
  n^{(3)}c(p_a^2+p_b^2)\nonumber\\
  & & +(ap_a+bp_b+cp_c)^2-{\textstyle\frac{1}{2}}(ap_a+bp_b)^2-(cp_c)^2+
  {\textstyle\frac{1}{2}}(ap_b-bp_a)^2  \label{hamlrs}
\end{eqnarray}
for LRS models, where the constants $n^{(I)}$ specify the Bianchi
type, and
\begin{equation} \label{hamisotropic}
  g_0^{-1}{\cal H}^{({\mathrm E})}= -2(n^{(1)}+n^{(2)}+n^{(3)})c p^2+6(c
  p)^2
\end{equation}
for isotropic models, where $n^{(1)}=n^{(2)}=n^{(3)}=0$ for the
isotropic flat model (isotropic Bianchi I), and
$n^{(1)}=n^{(2)}=n^{(3)}=1$ for the isotropic closed model (isotropic
Bianchi IX).

To check the constraint algebras, we will also need the diffeomorphism
constraints, which are
\begin{equation} \label{diffbianchi}
  {\cal D}_I=-(\kappa\iota')^{-1}c^K_{IJ}\phi^i_Kp^J_i
\end{equation}
for Bianchi models, and vanish identically for LRS models (provided
that the Gau\ss\ constraint is solved) and isotropic models. So only
for Bianchi models demanding an anomaly free representation of the
constraint algebra will put restrictions on the quantization of the
constraints, although not for all types (for Bianchi type I the
structure constants vanish, and so the diffeomorphism constraint).

For a non-vanishing diffeomorphism constraint, the constraint algebra
can be derived from the full theory: the Poisson bracket of a
diffeomorphism constraint smeared with a shift vector field $N^a$ and
a Hamiltonian constraint smeared with lapse function $N$ yields a
Hamiltonian constraint with a lapse function given by the shift vector
field and the derivatives of the lapse function $N$. In a reduction to
homogeneous configurations, the lapse function is a constant, and so
its derivatives vanish. Therefore, the diffeomorphism constraint and
the Hamiltonian constraint Poisson commute. Moreover, as can easily be
checked by a direct calculation using the Jacobi identity for the
structure constants $c^K_{IJ}$, the diffeomorphism constraint
(\ref{diffbianchi}) commutes with the Euclidean part of the
Hamiltonian constraint (\ref{hambianchi}).

We now come to a description of the kinematical sector of loop quantum
cosmology \cite{cosmoI}. For Bianchi models, the three scalars
$\phi^i_I\tau_i$ are exponentiated to point holonomies
\cite{FermionHiggs} taking values in $SU(2)$. The auxiliary Hilbert
space $\Haux=L^2(SU(2)^3,\md\mu_{\mathrm H}^3)$ consists of functions
of these point holonomies being square integrable with respect to the
three-fold copy of Haar measure. A basis is given by spin network
states which combinatorially can be associated to spin networks with
three closed edges meeting in a single $6$-vertex $x_0$. After
extending $x_0$ to an auxiliary manifold, they can be identified with
spin networks embedded in this manifold. The diffeomorphism constraint
generates transformations which correspond to conjugations in $S$,
e.g.\ rotations for the Bianchi type IX model.

The auxiliary Hilbert spaces of LRS and isotropic models can be
obtained from that of Bianchi models by a further reduction which
takes care of the fact that for those models point holonomies which
are rotated by an element of the isotropy group are gauge equivalent:
while kinematical quantum states for Bianchi models are given by
functions $f(h_1,h_2,h_3)$ with $h_I:=\exp(\phi^i_I\tau_i)\in SU(2)$,
for LRS models quantum states are given by functions
\[
 f(h_1,\exp({\ts\frac{\pi}{2}}\Lambda_3) h_1
 \exp(-{\ts\frac{\pi}{2}}\Lambda_3), h_3)
\]
where $h_3$ is a function of $\Lambda_3:=\Lambda_3^i\tau_i$ via
$h_3=\exp(c\Lambda_3)$, and for isotropic models by functions
\[
  f(\exp({\ts\frac{\pi}{2}}\Lambda_2) h_3 \exp(-{\ts\frac{\pi}{2}}\Lambda_2),
  \exp(-{\ts\frac{\pi}{2}}\Lambda_1) h_3 \exp({\ts\frac{\pi}{2}}\Lambda_1),
  h_3)\,.
\]
Therefore, their states can be represented with a reduced number of
edges. However, the configuration spaces are not subgroups of
$SU(2)^3$, but only subsets given by a union of conjugacy classes.
This implies that quantum states can no longer be represented as
ordinary spin network states, but as generalized spin network states
with insertions (see Ref.\ \cite{cosmoII} for a discussion).

\section{The Hamiltonian Constraint in the Full Theory}
\label{s:Full}

For real Ashtekar variables, the Hamiltonian constraint in Lorentzian
signature is given by \cite{AshVarReell,QSDI}
\begin{equation} \label{Ham}
   {\cal H}[N]  = \int_{\Sigma}\md^3x N(\det q)^{-\frac{1}{2}}
 \epsilon^{ij}\mbox{}_k E_i^aE_j^b\left(F_{ab}^k-2(1+\iota^2)E^a_{[i}
 E^b_{j]} K^i_aK^j_b\right)
\end{equation}
where $F_{ab}$ are the curvature components of the Ashtekar connection
and
\[
  K^i_a:=(\det q)^{-\frac{1}{2}}K_{ab}E^{bi}
\]
are the coefficients of the extrinsic curvature of $\Sigma$. The
extrinsic curvature coefficients are complicated functions of the
phase space variables, but luckily they are contained only in the
second term of the constraint (\ref{Ham}). An important step towards a
quantization of the Hamiltonian constraint has been done in Refs.\ 
\cite{QSDI,QSDII} leading to a first consistent regularization of the
constraint. An important input was a procedure to treat the second
term of the constraint containing the curvature coefficients $K_a^i$,
which has been achieved by first writing
\begin{equation}\label{HamLor}
 {\cal H}=2(\det q)^{-\frac{1}{2}}(1+\iota^2)\tr([K_a,K_b]
 [E^a,E^b])-\HE
\end{equation}
where
\begin{equation}\label{Eucl}
 \HE=2(\det q)^{-\frac{1}{2}}\tr(F_{ab}[E^a,E^b])
\end{equation}
is the Euclidean constraint. In these formulae the variables are
written as ${\cal L}SU(2)$-valued, e.g.\ $E^a:=E^a_i\tau^i$. The
square brackets and $\tr$ denote the commutator and trace in this Lie
algebra.

It is then necessary to represent the curvature components
$K_a=K_a^i\tau_i$ as quantum operators. The recipe of Ref.\
\cite{QSDI} starts from the observation that the trace
\[
 K:=\int_{\Sigma}\md^3x\sqrt{\det
  q}K_{ab}q^{ab}=\int_{\Sigma}\md^3xK_a^iE_i^a
\]
is the time derivative of the volume (independent of the space-time
signature) which can in a Hamiltonian formulation be written as
Poisson bracket of Hamiltonian and volume:
\[
 K={\cal L}_tV=-\{V,\HE\}\,.
\]
When one then uses the important fact that $\{\Gamma^i_a,K\}=0$
\cite{Reality} to represent the components of the intrinsic curvature as
\[
 K_a^i=\frac{\delta K}{\delta E^a_i}=(\kappa\iota)^{-1}\{A_a^i,K\}\,,
\]
translates all Poisson brackets into commutators, expresses the
connection coefficients by holonomies and uses a quantization of the
volume \cite{AreaVol,Vol2,Vol} one can trace back the quantization of the
second term of (\ref{Ham}) to a quantization of the first term.

Furthermore, in Ref.\ \cite{QSDI} a quantization of the Euclidean
constraint $\HE$ has been given which completes the quantization of
${\cal H}$. Here a key point is that in a diffeomorphism invariant
context only expressions with the correct density weight can be
quantized without a background structure. This means that the
determinant of the metric must not be absorbed into the lapse
function in order to have a polynomial constraint because this would
change the density weight. Instead, the key identity
\begin{equation} \label{key}
 (\det q)^{-\frac{1}{2}}[E^a,E^b]^i=\epsilon^{abc}e_c^i=
 2\epsilon^{abc}\frac{\delta V}{\delta E^c_i}=2(\kappa\iota)^{-1}
 \epsilon^{abc}\{A_c^i,V\}
\end{equation}
is used and quantized by turning the Poisson bracket into a
commutator, representing the connection coefficients by a holonomy and
using the known volume operator $\hat{V}$.

To regularize the constraint, the manifold $\Sigma$ is triangulated in
a way adapted to the graph of a cylindrical function on which the
constraint operator is to act. This is done in such a manner that for
each triple of edges meeting in a vertex $v$ a triangle $\alpha_{ij}$,
which consists of two pieces $s_i$ and $s_j$ of two edges and a third
curve $a_{ij}$ connecting their endpoints, is chosen as a base of the
tetrahedron spanned by $s_i$, $s_j$ and a piece $s_k$ of the third
edge. The rest of $\Sigma$ is triangulated arbitrarily. Using the
above identities and an expansion of holonomies for infinitesimal
length one can see that the vertex contributions
\[
 \epsilon^{ijk}\tr(h_{\alpha_{ij}}h_{s_k}[h_{s_k}^{-1},\hat{V}])
\]
summed over all triples of edges meeting in $v$ have the correct
classical limit for a local contribution of the Euclidean constraint.

Contained in the classical limit is a continuum limit in which the
width of the triangulation vanishes. In the quantum theory, however,
such a continuum limit is trivial provided we quantize the constraint
on the Hilbert space $\Hdiff$ where the diffeomorphism constraint is
solved. In this case two operators which are obtained by a
triangulation and a refinement of it are identical because their
actions on a fixed cylindrical function by construction differ only by
a diffeomorphism moving the edges $a_{ij}$.

When acting on a cylindrical function the Euclidean constraint changes
the graph due to the holonomies to the curves $a_{ij}$ in such a way
that in the neighborhood of any non-planar vertex for each two edges
incident there a new edge connecting them is generated. The action of
the operator is well understood, but there are only some rather
trivial explicitly known solutions and the complete solution space is
not under any control. Furthermore, there are a lot of ambiguities in
the construction of a particular operator and it is not clear which
one to use. This question can probably be answered only by
investigating whether the theory has the correct classical limit. A
strategy proposed here is to use similar regularizations of the
Hamiltonian constraint on appropriate sectors of symmetric states for
a comparison with the corresponding classical mini-superspaces.

\section{Hamiltonian Constraints for Bianchi Models}
\label{s:Bianchi}

We are now ready to present a quantization of the Hamiltonian
constraints for cosmological models, starting with Bianchi models. Our
construction will follow that of Ref.\ \cite{QSDI} in the full
theory, but there are important points where new input is necessary.
This is the case because, first, our regularization has to respect the
symmetry conditions and, second, because connection coefficients are
coded in point holonomies rather than ordinary holonomies.
Furthermore, Eq.\ (\ref{hambianchi}) shows that the constraint
contains two different parts: a part linear in the scalar fields which
via the structure constants $c^K_{IJ}$ depends on the Bianchi type,
and a second part quadratic in the scalar fields which is independent
of the type. Therefore, although the original expression (\ref{Ham})
does not bear any reference to the symmetry type, all the reduced
models will have different Wheeler--DeWitt operators and different
dynamics.

As already mentioned, we will follow the procedure presented in Ref.\ 
\cite{QSDI} and recalled in Section \ref{s:Full} when quantizing the
expressions (\ref{hambianchi}), (\ref{hamlrs}) and
(\ref{hamisotropic}) on the respective kinematical Hilbert spaces.
Some steps can immediately be copied, for instance we can restrict
ourselves to the Euclidean constraints because the additional part in
the Lorentzian constraint can be treated in complete analogy to the
full theory by using the extrinsic curvature $K$. Furthermore, we will
also make use of commutators of holonomies with the volume operator,
i.e.\ this operator will again play a prominent role. At this place it
is fortunate that it simplifies in symmetric regimes \cite{cosmoII}
such that an analysis of the matrix elements of the Hamiltonian
constraint will be easier.

What we cannot copy in our reduced models is the use of a
triangulation of space and the subsequent continuum limit. At first
sight, we may seem to be in a better situation because we are
considering a finite dimensional model and do not have to bother about
a continuum limit. But in the regularization used in the full theory
this limit also serves the purpose to express the curvature components
$F^i_{ab}$ entering (\ref{Ham}) in terms of closed loops in the
following way: As recalled above, one starts with a triangulation of
space which is adapted to the graph of a spin network the Hamiltonian
constraint operator is supposed to act on. Forming a triangular loop
$\alpha_{ij}$ based in $v$ one can compute the holonomy
$h_{\alpha_{ij}}$ appearing in each vertex contribution discussed in
Section \ref{s:Full}. In the continuum limit any such loop shrinks to
$v$ such that the holonomies approximate the curvature components.
Here one makes use of the expansion
\[
 h(A)=1+\half \epsilon^2u^av^bF_{ab}^i\tau_i+O(\epsilon^3)
\]
for a holonomy $h$ along an infinitesimal parallelogram spanned by two
edges of length $\epsilon$ with directions $u^a$ and $v^a$ in
Euclidean space. This is the final ingredient needed to turn all
contributions to the classical Hamiltonian constraint into quantum
operators.

The ensuing Wheeler--DeWitt operator then contains the holonomies
along all the triangular loops $\alpha_{ij}$ as multiplication
operators creating new edges $a_{ij}$. Although the construction of
these loops depends on the triangulation, the action of the operator
is triangulation independent once it is considered to act on the
kinematical Hilbert space where the diffeomorphism constraint is
solved. In this sense, the continuum limit is trivial in the quantum
theory because for a finer triangulation the newly created edges are
just moved using a diffeomorphism thus representing the same
diffeomorphism invariant quantum state.

Obviously, we cannot use these techniques to generate the curvature
components in a symmetric model: First, the prescription to generate
new edges violates our symmetry conditions (e.g., in the Bianchi
models there are just the three edges obtained by smearing the point
holonomies, and no edge connecting two of them can be generated).
Second, we do not have a continuum limit in which course loops would
shrink and could be approximated using curvature components.

We will present now alternative techniques which are adapted to the
symmetry and which make use of the methods developed to deal with
point holonomies. As in the computation of the derivative operators in
Ref.\ \cite{cosmoI} the regularizing edges will be important. All this
will be discussed in detail for the Bianchi models and then used in
LRS and isotropic models.

As demonstrated in Ref.\ \cite{QSDI}, we have to provide the Euclidean
part (\ref{hambianchi}) of the Hamiltonian constraint with the correct
density weight in order to be able to quantize it in a background
independent manner. We thus divide the earlier expression by
$\sqrt{\det q}= \sqrt{\frac{1}{6}g_0 |\epsilon^{ijk} \epsilon_{IJK}
  p_i^Ip_j^Jp_k^K|}$, multiply with a constant (due to homogeneity)
lapse function $N$ and integrate over the manifold $\Sigma$:
\[
 \HE[N]= -\int_{\Sigma}\md^3 x N
  (\det q)^{-\frac{1}{2}} g_0\epsilon^{ijk}p_i^Ip_j^JF_{IJk}
\]
with
\[
  F_{IJ}^i=-c^K_{IJ}\phi^i_K+\epsilon^i{}_{jk}\phi^j_I\phi^k_J\,.
\]
This can immediately be integrated to
\begin{eqnarray*}
 \HE[N] & = & -V_0N\left({\textstyle\frac{1}{6}}
   \left| \epsilon^{lmn}\epsilon_{LMN}p_l^Lp_m^Mp_n^N\right|
 \right)^{-\frac{1}{2}} \epsilon^{ijk}p_i^Ip_j^JF_{IJk} \\
 & = & -{\textstyle\frac{1}{6}}V_0(\kappa\iota')^{-1}N \left(
   {\textstyle\frac{1}{6}} \left| \epsilon^{lmn} \epsilon_{LMN}
     p_l^Lp_m^Mp_n^N \right|\right)^{-\frac{1}{2}} \left\{\phi_K^k,
   \epsilon_{OPQ} \epsilon^{ijl}p_i^Op_j^Pp_l^Q\right\} \epsilon^{IJK}
   F_{IJk}\\
 & = & -2(\kappa\iota')^{-1} N\epsilon^{IJK} \left\{\phi_K^k,V
   \right\} F_{IJk}
\end{eqnarray*}
where we used
\[
 V=V_0\sqrt{{\ts\frac{1}{6}}|\epsilon^{ijk}\epsilon_{IJK}
   p^I_ip^J_jp^K_k|}
\]
and
\begin{equation}\label{commu}
  \left\{\phi_K^k,\epsilon_{MNL}\epsilon^{ijl}p_i^Mp_j^Np_l^L\right\}=
  3 \kappa\iota' \epsilon^{ijk} \epsilon_{MNK} p_i^Mp_j^N
\end{equation}
which, divided by $\sqrt{\det q}$ is essentially the key identity
(\ref{key}) used in Ref.\ \cite{QSDI} to quantize the inverted triad
components. The expression of $F^k_{IJ}$ in terms of the scalar fields
depends on the particular symmetry group, i.e.\ on the Bianchi type,
which will be taken care of when we choose the route of loops below.

\subsection{Auxiliary Manifolds}

But before this we have to discuss our regularization scheme which, as
already noted, has to be different from that in the full theory
because we cannot use a triangulation as regulator and the continuum
limit in connection with diffeomorphism invariance to remove it. After
integrating over $\Sigma$ we arrived above at an expression for the
Hamiltonian constraint which is completely composed of the reduced
field components sitting in the single point $x_0$ of the reduced
manifold $B:=\Sigma/S=\{x_0\}$. Obviously in this reduced description
there is no substitute for a triangulation of space which could be
used as regulator. Note, however, that in order to compute derivative
operators in point holonomies \cite{cosmoI} we already had to smear
point holonomies along auxiliary edges. To that end we have, in the
present context of cosmological models, introduced an auxiliary
manifold $\overline{S/F}$ (a suitable compactification of the
homogeneous space $S/F$) in which these edges are to lie, where in the
construction of both the auxiliary manifold and the edges we made use
of the structure of $\overline{S/F}$ as a homogeneous space. This
structure can now also be used in order to regulate the Hamiltonian
constraint, but again we cannot simply copy the procedure of the full
theory (applied to the $6$-vertex lying in the auxiliary manifold)
because this would spoil the symmetry (we are not allowed to create
new edges, but only to retrace existing ones).

Although we then manage to respect the symmetry, we immediately have
to face another problem: In order to approximate the curvature
component by the attached loops and to recover the correct classical
limit we have to shrink the loops to the vertex, which in the full
theory was achieved by the continuum limit which we now do not have at
our disposal. Our only possibility is to shrink the whole auxiliary
manifold and with it the auxiliary edges to the point $x_0$. In this
limit we will recover the classical limit, whereas the quantum theory
is independent of the extension of our auxiliary manifold. This is
completely analogous to the full theory, where the diffeomorphism
invariant quantum theory is triangulation independent.

Let us now follow in detail the program outlined above. We have a
compact homogeneous auxiliary manifold $\overline{S/F}$ containing the
point $x_0$. Each point of $\overline{S/F}$ can be reached from $x_0$
by following pieces of integral curves to the left invariant vector
fields $X_I$ (which also have been used to define the auxiliary edges
smearing point holonomies). If we always normalize the invariant
vector fields in such a way that their closed (by construction of
$\overline{S/F}$) integral curves have a fixed parameter length, then
shrinking the auxiliary manifold is equivalent to multiplying the
vector field by a number $\epsilon$ smaller than one. So we
can describe the shrinking of $\overline{S/F}$ to $\SFe$ by replacing
$X_I$ with $\epsilon X_I$ where eventually we will consider the limit
$\epsilon\to 0$ in connection with the classical limit.

We now illustrate the shrinking procedure for the Bianchi I model on
the manifold $\R^3$. In terms of coordinates $x^I$ adapted to the
symmetry (the usual Cartesian coordinates), left invariant vector
fields are $X_I=\frac{\partial}{\partial x^I}$. The manifold $\R^3$ is
non-compact, but we can compactify it to a three-torus, which then
plays the role of $\SF$, by restricting the coordinates to $0\leq
x^I\leq 1$ and identifying the points with $x^I=0$ and $x^I=1$ for
each $I$. By this compactification the integral curves of the vector
fields $X_I$ are rendered closed and will be used as auxiliary edges
based in a base point $x_0$ in the three-torus. Denoting the flow
generated by a vector field $X$ by $\Phi_t(X)\colon\SF\to\SF$, we 
can write
\[
  \SF=\{\Phi_{t_3}(X_3)\Phi_{t_2}(X_2)\Phi_{t_1}(X_1)x_0:
  t_I\in S^1\cong\R/\Z\}\,.
\]
The shrinking manifolds $\SFe$ are then obtained for fixed
$t_I$-intervals by replacing $X_I$ with $\epsilon X_I$ everywhere.
Then the three-torus as well as the closed integral curves shrink and,
for infinitesimal $\epsilon$, holonomies along the integral curves
(i.e.\ the smeared point holonomies) are approximated by $\epsilon$
times the respective scalar field component
($h_I=1+\epsilon\phi_I+O(\epsilon^2)$), and volume integrals over the
auxiliary manifold are approximated by $\epsilon^3$ times the reduced
quantity defined in the base point.

\subsection{Regularization}

Our previous expression for the Hamiltonian constraint was written
down in the reduced description on the point $x_0$. To extend it to
the auxiliary manifold we just have to integrate it, yielding just a
factor of $\epsilon^3$ because the constraint is constant on $\SFe$.
We thus have
\begin{equation}\label{HamBianchiReg}
 \HE_{\epsilon}[N]:= \int_{\SFe}\md^3x \HE[N]=
  -2\epsilon^3 (\kappa\iota')^{-1} N\epsilon^{IJK}
  \left\{\phi_K^k,V \right\} F_{IJk}
\end{equation}
as point of departure for the quantization. An integration is
mandatory here in order to obtain an expression without density weight
(which can be defined using scale transformations respecting the
symmetry). The original, unregulated expression can be recovered as
(in this expression the limit is trivial which will, however, not be
the case if we introduce holonomy variables below)
\begin{equation}\label{HamReg}
  \lim_{\epsilon\to 0}\epsilon^{-3}\int_{\SFe}
  \md^3x \HE[N]=\HE[N]\,.
\end{equation}
One factor of $\epsilon$ can be used to express $\epsilon\phi_K^k$ as
infinitesimal holonomy, whereas the remaining square of $\epsilon$
will be used to express the curvature components via a holonomy along
an infinitesimal loop, to which we turn now.

Because the curvature components, expressed in terms of scalar fields,
are the only place where the Hamiltonian constraint depends on the
symmetry group, we have to make use of its structure when expressing
the curvature by means of suitably laid loops. This can be done most
easily by composing the loop of pieces of integral curves to the
invariant vector fields, i.e.\ as a geodesic parallelogram in an
arbitrary homogeneous metric. E.g., in the Bianchi I model we can use
a chart with coordinates $x^I$ which is adapted to the symmetry in
such a way that the (commuting) invariant vector fields are given by
$X_I=\frac{\partial}{\partial x^I}$. Using the flows $\Phi_t(X_I)$
generated by these vector fields, we can form parallelograms in $\SF$
composed of the four edges $c_1\colon t\mapsto \Phi_t(X_I)x_0$,
$c_2\colon t\mapsto \Phi_t(X_J)c_1(1)$, $c_3\colon t\mapsto
\Phi_{-t}(X_I)c_2(1)$ and $c_4\colon t\mapsto \Phi_{-t}(X_J)c_3(1)$,
for all $1\leq I\not=J\leq 3$, which is closed because the vector
fields $X_I$ and $X_J$ commute. We can now use the constant scalar fields
$\phi_I=\phi(X_I)$ parameterizing a homogeneous connection
$A^i_a=\phi^i_K\omega^K_a$ to compute
the holonomy along the parallelogram: $\exp(\phi_I) \exp(\phi_J)
\exp(-\phi_I) \exp(-\phi_J)$ using, e.g.\
\[
  h_{c_1}(A)={\cal P}\exp\int_{c_1}\md t\,\dot{e}^aA_a^i\tau_i= {\cal
    P}\exp\int_{c_1}\md t\, X_I^a\omega^K_a\phi^i_K\tau_i=
  \exp(\phi_I^i\tau_i)\,.
\]

What remains is to note that point holonomies were defined in just the
same way by calculating holonomies along auxiliary edges defined as
integral curves to an invariant vector field \cite{cosmoI}. We only
have to take care that we run through a complete closed integral curve
to a single invariant vector field which is possible due to our
construction of the compactification $\SF$. There are then two
possibilities to shrink the parallelogram: First we can fix the
homogeneous manifold and only shrink the curves by constraining the
parameter $t$ above to an interval $[0,\epsilon]$. This is the usual
procedure for a regularization in a three-dimensional theory, but is
not applicable here because we can identify a holonomy along an
integral curve with a point holonomy only if the curve is closed and
based in the point $x_0$. Thus we have to use the second shrinking
procedure where both the auxiliary manifold and the curves embedded
into it shrink by using the vector fields $\epsilon X_I$ as generators
of both the auxiliary manifold $\SFe$ and of their own closed integral
curves. Now the holonomies are always along closed integral curves and
can be identified with point holonomies for all values of $\epsilon$,
but due to $\phi(\epsilon X_I)=\epsilon\phi_I$ (linearity of the map
$\phi\colon{\cal L}N\to{\cal L}G$) we now have the
$\epsilon$-dependent holonomy
\begin{eqnarray*}
 && \exp(\epsilon\phi_I) \exp(\epsilon\phi_J) \exp(-\epsilon\phi_I) 
    \exp(-\epsilon\phi_J) \\
 && \quad = (1+\epsilon\phi_I+{\ts\frac{1}{2}}\epsilon^2\phi_I^2)
 (1+\epsilon\phi_J+{\ts\frac{1}{2}}\epsilon^2\phi_J^2)
 (1-\epsilon\phi_I+{\ts\frac{1}{2}}\epsilon^2\phi_I^2)
 (1-\epsilon\phi_J+{\ts\frac{1}{2}}\epsilon^2\phi_J^2)
 +O(\epsilon^3)\\
 && \quad = 1+\epsilon^2(\phi_I\phi_J-\phi_J\phi_I)+O(\epsilon^3)\\
  && \quad = 1+\epsilon^2\epsilon_{ijk}\phi_I^i\phi_J^j\tau^k+
 O(\epsilon^3)= 1+\epsilon^2F_{IJ}^k\tau_k+O(\epsilon^3)\,.
\end{eqnarray*}
Thus we have a product of point holonomies which,
for infinitesimal $\epsilon$, reproduces the curvature components for
Bianchi I.

For other Bianchi models with a non-Abelian symmetry group, the last
equation does not recover the correct curvature components because of
the additional term containing the structure constants $c^K_{IJ}$ of
the symmetry group. But, also due to the non-Abelian nature of the
symmetry group, the parallelogram constructed above will not be
closed, even up to the order of $\epsilon^2$ because now the vector
fields used to define the edges of the parallelogram do not commute.
As is well-known from differential geometry, the endpoint $c_4(1)$ of
the above parallelogram constructed from the vector fields $\epsilon
X_I$ and $\epsilon X_J$ is a distance $[\epsilon X_I,\epsilon X_J]$
away from the origin. So we can close our parallelogram by running
through the curve $c_5\colon t\mapsto\Phi_{-t}([\epsilon X_I,\epsilon
X_J])c_4(1)$ after following the curves $c_1$ to $c_4$. The ensuing
curve is closed to the order of $\epsilon^3$ and can be closed by
connecting the points $c_5(1)$ and $x_0$ which affects the holonomy
along the loop only up to the same order. As holonomy we obtain
\begin{eqnarray*}
 \lefteqn{\exp(\epsilon\phi_I) \exp(\epsilon\phi_J) \exp(-\epsilon\phi_I) 
    \exp(-\epsilon\phi_J) \exp(-\epsilon^2\phi([X_I,X_J]))}\\
 & = &\exp(\epsilon\phi_I) \exp(\epsilon\phi_J) \exp(-\epsilon\phi_I)
 \exp(-\epsilon\phi_J) \exp(-\epsilon^2c^K_{IJ}\phi_K)
\end{eqnarray*}
which, thanks to the commutator of left invariant vector fields,
depends explicitly on the structure of the symmetry group. Note also
that the commutator of two vector fields results in a factor of
$\epsilon^2$ for only one power of $\phi$ in the last exponential. Using the
previous calculations we can easily convince ourselves that this leads
to the correct curvature components:
\begin{eqnarray*}
 \lefteqn{\exp(\epsilon\phi_I) \exp(\epsilon\phi_J) \exp(-\epsilon\phi_I) 
    \exp(-\epsilon\phi_J) \exp(-\phi([\epsilon X_I,\epsilon X_J]))}\\
 & = & (1+\epsilon^2\epsilon_{ijk}\phi_I^i\phi_J^j\tau^k+
 O(\epsilon^3))(1-\epsilon^2c^K_{IJ}\phi_K+O(\epsilon^4))\\
 & = & 1+ \epsilon^2 (-c^K_{IJ}\phi_K^k+ \epsilon_{ijk}
 \phi_I^i\phi_J^j) \tau_k+O(\epsilon^3)= 1+ \epsilon^2 F^k_{IJ}
 \tau_k+ O(\epsilon^3)\,.
\end{eqnarray*}

We are now in a position to put all ingredients together in order to
arrive at our regulated classical expression of the Hamiltonian
constraint formulated on the regularizing manifold $\SFe$. Comparing
with Eq.\ (\ref{HamBianchiReg}), we see that
\begin{eqnarray*}
 \HE_{\epsilon}[N] & := & 4 (\kappa\iota')^{-1} N
  \sum_{IJK}\epsilon^{IJK} \tr( \exp(\epsilon\phi_I)
   \exp(\epsilon\phi_J) \exp(-\epsilon\phi_I) \exp(-\epsilon\phi_J)\\
 & & \times \exp(-\epsilon^2\phi([X_I,X_J])) \{\exp\epsilon\phi_K,V\})
\end{eqnarray*}
has the correct limit (see Eq.\ (\ref{HamReg}))
\[
 \HE[N]=\lim_{\epsilon\to 0}\epsilon^{-3}\HE_{\epsilon}[N]
\]
when the regulator is removed.

\subsection{Quantization}

From now on we can again follow the steps of Ref.\ \cite{QSDI} in
order to quantize the constraint: exponentials of scalars are replaced
by (point) holonomies, the volume is quantized to the operator of
Ref.\ \cite{cosmoII} and the Poisson bracket is replaced with
$(i\hbar)^{-1}$ times a commutator. The result is
\begin{equation}\label{HamEuclBianchiQuant}
 \hatHE[N]=-4i(\iota'\lP^2)^{-1} N \sum_{IJK}\epsilon^{IJK}
 \tr\left( h_Ih_Jh_I^{-1}h_J^{-1} h_{[I,J]}^{-1} [h_K,\hat{V}]\right)
\end{equation}
where $h_I$ is the holonomy along the $I$-th regularizing edge,
interpreted as multiplication operator, and $h_{[I,J]}$ depends on the
symmetry group in the following way
\[
 h_{[I,J]}:=\prod_{K=1}^3 (h_K)^{c^K_{IJ}}\,.
\]

What remains to do is to use the decomposition (\ref{HamLor}) and the
quantization of the Euclidean part of the constraint in order to
quantize the Lorentzian constraint. In the reduced formulation the
decomposition reads (for a constant lapse function $N$)
\begin{eqnarray*}
  {\cal H}[N] & = & 2(1+\iota^2)\int_{\Sigma}\md^3x N 
  (\det q)^{-\frac{1}{2}} \tr([K_a,K_b] [E^a,E^b])-\HE[N]\\
  & = & 2(1+\iota^2)\int_{\Sigma}\md^3x N \sqrt{g_0} 
  \left|\det(p_K^l)\right|^{-\frac{1}{2}}
  \tr( [k_I,k_J][p^I,p^J])-\HE[N]\\
  & = & -(1+\iota^2) V_0N\epsilon_{ijk}k_I^ik_J^j
  \left|\det(p_K^l)\right|^{-\frac{1}{2}}
  \epsilon^{klm} p^I_lp^J_m-\HE[N] \\
  & = & -2(1+\iota^2) (\kappa\iota')^{-3} V_0^{-2} N
  \epsilon_{ijk} \epsilon^{IJK} \{\phi_I^i,K\} \{\phi_J^j,K\}
  \{\phi_K^k,V\}-\HE[N]
\end{eqnarray*}
where we defined $K_a^i=:k_I^i\omega^I_a$, $K=\int\md^3x K_a^iE^a_i=
V_0k_I^ip^I_i$ and used Eq.\ (\ref{commu}). Regularized as above, we
obtain
\begin{eqnarray*}
  {\cal H}_{\epsilon}[N] & = & -2(1+\iota^2) (\kappa\iota)^{-3} V_0 N
  \epsilon_{ijk} \epsilon^{IJK} \epsilon^3 \{\phi_I^i,K\} \{\phi_J^j,K\}
  \{\phi_K^k,V\}-\HE_{\epsilon}[N]\\
  & = & 8(1+\iota^2) (\kappa\iota)^{-3} V_0 N \epsilon^{IJK}
  \tr( \{\exp(\epsilon\phi_I),K\} \{\exp(\epsilon\phi_J),K\}
  \{\exp(\epsilon\phi_K),V\})\\
  && -\HE_{\epsilon}[N]+O(\epsilon^3)
\end{eqnarray*}
which can again be quantized by replacing the exponentials of scalars
by (point) holonomies and the Poisson brackets by $(i\hbar)^{-1}$
times a commutator:
\begin{equation}\label{HamBianchiQuant}
 \hat{{\cal H}}[N]=8i(1+\iota^2)(\iota\lP^2)^{-3}V_0 N
 \epsilon^{IJK} \tr\left([h_I,\hat{K}] [h_J,\hat{K}] [h_K,\hat{V}]\right)-
 \hatHE[N]\,.
\end{equation}
As in the full theory, we use here the quantization
\[
  \hat{K}=i\hbar^{-1}\left[\hat{V},\hatHE[1]\right]
\]
of the integrated extrinsic curvature.

A non-vanishing cosmological constant $\Lambda$ can be included simply
by adding the contribution $2N\Lambda\hat{V}$ using the volume operator
of Ref.\ \cite{cosmoII}.

Of course, there are factor ordering ambiguities which we ignored in
writing the expression above. Furthermore, as written down, the
operator is not symmetric which would have to be achieved by choosing
an appropriate factor ordering. In the full theory these issues are
unsolved and one purpose of the reduced models discussed here can be
to gain insights by studying them in a simplified, highly symmetric
regime. We will, however, not enter this discussion because it
requires a detailed study of these models, whereas here we are mainly
interested in general aspects.

It is immediate to see that the operator (\ref{HamBianchiQuant}) is
manifestly independent of the regulator $\epsilon$ which was used to
reformulate the classical expression in terms of holonomy variables.
This is so because, in contrast to the classical procedure where the
scalar fields are fixed in the limit $\epsilon\to 0$ in which the
regulator is removed, in the quantum theory the scalars diverge in
this point limit in such a way that smeared holonomies are equal to
point holonomies independent of $\epsilon$. Recall that the quantum
configuration space contains distributional scalars which are needed
to accomplish this result. Therefore, it does not matter whether we
formulate the operator on the auxiliary manifold $\SF$, on one of the
shrunk manifolds $\SFe$ or even in the point limit using exclusively
point holonomies. In each of these formulations we have the same state
space and the same action of the Hamiltonian constraint operator. This
feature, which here arose after smearing the scalar fields to regulate
classical expressions, substitutes the diffeomorphism invariance which
in the full theory is used to show that the continuum limit removing
the regulator there is trivial in the quantum theory. Thus, although
formulated in a conceptually different way, the quantization of the
Hamiltonian constraint in the full theory \cite{QSDI} and ours for the
Bianchi models share the same key properties.  But of course our
discussion was technically simplified by the fact that we only have to
regard $6$-vertices, whereas in the full theory there can be (and have
to be taken into account) vertices of arbitrary valence.

It is immediate to see that the constraints are represented in an
anomaly free manner: the only non-trivial part is to check that
diffeomorphisms commute with the Hamiltonian constraint which can
easily be seen after recalling that diffeomorphisms, i.e.\ inner
automorphisms on the auxiliary manifold, just move the three edges of
homogeneous spin networks, whereas the Hamiltonian constraint operator
fixes them and only changes their labels and the contractor.

\section{Hamiltonian Constraints for LRS and\\ Isotropic Models}
\label{s:Iso}

Similarly to the treatment of volume operators in Ref.\ \cite{cosmoII}
we can use the expression (\ref{HamBianchiQuant}) and insert the
rotated holonomies $h_2=\exp(\frac{\pi}{2}\Lambda_3)h_1
\exp(-\frac{\pi}{2}\Lambda_3)$ to arrive at the constraint operator
for LRS models, and $h_1=\exp(\frac{\pi}{2}\Lambda_2)h_3
\exp(-\frac{\pi}{2}\Lambda_2)$, $h_2=\exp(-\frac{\pi}{2}\Lambda_1)h_3
\exp(\frac{\pi}{2}\Lambda_1)$ to arrive at the operator for isotropic
models. The resulting operator then only contains the holonomy
operators $h_1$, $h_3$ for LRS models and $h_3$ for isotropic models,
and in addition operators coming from the exponentials of $\Lambda_I$
which manipulate the insertions (see Ref.\ \cite{cosmoII}).

By construction, all these operators are gauge invariant, but the
intermediate action of holonomies and the volume operator is on gauge
non-invariant states. This means that the techniques used for gauge
invariant states of isotropic models in Ref.\ \cite{cosmoII} have to
be generalized so as to deal with non-invariant states. In particular,
the volume operator derived there is not sufficient to calculate the
matrix elements of the Hamiltonian constraint for isotropic models.

In contrast to Bianchi models, where the three scalars $\phi_I$ and so
the three point holonomies $h_I$ are independent, for LRS and
isotropic models there are restrictions for the point holonomies (see
Eq.\ (\ref{scalar})) leading to the rotated holonomies above. This
implies that we can apply the following lemma, which has already been
used in Ref.\ \cite{cosmoII}, in order to simplify the constraint
operator which contains a product of several holonomy operators.

\begin{lemma} \label{gh}
  Let $g:=\exp (A\tau_i)$ and $h:=\exp (B\tau_j)$ with
  $A,B\in\R$, $i\not=j$ be matrices in the fundamental representation of
  $SU(2)$. Then
 \[ 
   gh=hg+h^{-1}g+hg^{-1}-\tr(gh)\,.
 \]
\end{lemma}

Let us illustrate this for the flat isotropic model, which
is obtained by restricting the Bianchi I model to isotropy. We thus
start from the expression (\ref{HamEuclBianchiQuant}) with all structure
constants vanishing, and insert the rotated holonomies $h_1$ and $h_2$
in terms of $h_3$. Then two different holonomies $h_I$ and $h_J$,
$I\not=J$, appearing in the operator are always of a form such that
Lemma \ref{gh} applies and we can insert
\[
  h_Jh_I^{-1}= h_I^{-1}h_J+ h_Ih_J+ h_I^{-1}h_J^{-1}- \tr(h_I^{-1}h_J)
\]
simplifying the sum in Eq.\ (\ref{HamEuclBianchiQuant}) to
\[
  \sum_{IJK}\epsilon^{IJK}\left( \tr([h_K,\hat{V}])+ \tr(
  h_I^2 [h_K,\hat{V}])+ \tr(h_J^{-2}[h_K,\hat{V}])-\tr(h_I^{-1}h_J)
  \tr(h_I h_J^{-1}[h_K,\hat{V}])\right)\,.
\]
The first trace is symmetric in (in fact independent of) $I$ and $J$
and therefore cancels in presence of $\epsilon^{IJK}$, and the rest
can be written as
\[
 \sum_{IJK}\epsilon^{IJK}\left( \tr((h_I^2-h_I^{-2}) [h_K,\hat{V}])
 -\tr(h_I^{-1}h_J) \tr(h_I h_J^{-1}[h_K,\hat{V}])\right)\,.
\]

Up to now the explicit expression for the rotated holonomies $h_1$ and
$h_2$ in terms of $h_3$ have not been used, but only the fact that
Lemma \ref{gh} applies because of the special character of rotated
holonomies. To arrive at the final expression we have to insert the
explicit formulae for $h_1$, $h_2$ which leads to a product of
$h_3$-holonomies and exponentials of $\Lambda_1$, $\Lambda_2$. Here we
can again apply Lemma \ref{gh} in order to collect all factors of
$h_3$ into a single power of $h_3$ (except for the one appearing on
the right hand side of the volume operator when the commutator is
written out) and a holonomy independent factor coming from the
exponentials of $\Lambda_I$. Alternatively, we can right from the
beginning apply Lemma \ref{gh} to the definition of $h_1$, $h_2$:
\begin{eqnarray*}
  h_1 & = & \exp({\ts\frac{\pi}{2}}\Lambda_2) h_3
    \exp(-{\ts\frac{\pi}{2}}\Lambda_2)\\
  & = & \exp({\ts\frac{\pi}{2}}\Lambda_2)\left(
    \exp(-{\ts\frac{\pi}{2}}\Lambda_2)h_3+ 
    \exp({\ts\frac{\pi}{2}}\Lambda_2)h_3+
    \exp(-{\ts\frac{\pi}{2}}\Lambda_2)h_3^{-1}-
    \tr(\exp(-{\ts\frac{\pi}{2}}\Lambda_2)
    h_3)\right)\\ 
  & = & h_3+h_3^{-1}+ \exp(\pi\Lambda_2)h_3-
   \exp({\ts\frac{\pi}{2}}\Lambda_2) 
   \tr(\exp(-{\ts\frac{\pi}{2}}\Lambda_2)h_3)
\end{eqnarray*} 
and similarly for $h_2$ such that $h_3$ appears always at the right
hand side. After some final applications of the lemma we can collect
all the factors containing $\Lambda_I$ in a single operator which
manipulates the insertion, whereas the power of $h_3$ changes the spin
of isotropic spin networks. 

Once the techniques of isotropic spin networks are generalized to gauge
non-invariant states, the calculation of matrix elements of the
Hamiltonian constraint operator in isotropic models can be pursued.
However, although there is no new input required in addition to the
procedure outlined here, the actual calculation can be expected to be
quite cumbersome and we do not present any further details. These
computations are needed when one begins to study specific models and
are well suited to be done numerically. In LRS models the situation is
similar: the essential steps have been outlined, but more work had to
be done for specific models.

Because for LRS and isotropic models the diffeomorphism constraint
vanishes on gauge invariant states, the constraint algebra is
represented anomaly free in a trivial way.

\section{Discussion}

In this paper we have presented quantizations of Hamiltonian
constraints for some cosmological models in the framework of loop
quantum cosmology. The strategy was to be as close to the full theory
of loop quantum gravity as possible in order to be able to compare
with and to draw conclusions for the full theory. We have seen that
this is possible to a large extent, but with necessary additional
input taking care of the symmetry. In particular, homogeneous models
are formulated in a single point $x_0$ and there is no continuum limit
which in combination with diffeomorphism invariance plays an important
role in the regularization of the Hamiltonian constraint in the full
theory. We have seen that an extension of the point $x_0$ to an
auxiliary manifold, which already appeared in the regularization of
derivative operators, provides the additional structure. The
Hamiltonian constraint is regularized on the auxiliary manifold, the
extension of which serves as regulator, i.e.\ the classical expression
is recovered if the auxiliary manifold is shrunk to the point $x_0$.
But the quantization of the regulated expression is independent of the
extension, and so we have an exact analog of the situation in the full
theory, where the continuum limit removes the regulator classically
and the quantization is regulator independent owing to diffeomorphism
invariance.

Whereas the investigation of homogeneous models in loop quantum
cosmology will be more complicated than and very different from the
standard treatment of minisuperspace quantizations, our constraint
operators are, compared with the Hamiltonian constraint operator of
the full theory, very similar but slightly simpler due to the
symmetry.  One simplification comes from the special nature of
vertices being at most $6$-valent. Also, the action is simpler because
the graph of a spin network being acted on is not changed: no new
edges are created but only the spins of existing ones and the vertex
contractor are changed.  Note that this implies that the large class
of special solutions to the Hamiltonian constraint found in Ref.\ 
\cite{QSDIII} has no counterpart in our cosmological models.
Furthermore, the fact that the constraint algebra is represented
anomaly free is realized rather trivially for the homogeneous models
because the classical algebra is simpler, and so requiring an anomaly
free representation is less restrictive as compared to the full theory
\cite{QSDIII,LM:Vertsm,Consist}.  The calculation of matrix elements
of the constraints for Bianchi models can be done along the lines of
Ref.\ \cite{Matrix}, whereas the computation for LRS or isotropic
models would be different because techniques for generalized spin
networks are needed there.

Another difference to the full theory is that we do not have to face
the issue of extracting a smooth, semiclassical metric from our
states. In the full theory, the quantum states represent a
distributional metric with a discrete structure which, in order to
perform a classical limit, have to be superposed in some way to
semiclassical states. In a homogeneous regime, however, the metric is
necessarily smooth and each state represents a homogeneous (but
nevertheless quantum) metric. But we still have to understand how to
interpret possible solutions of the constraint in a space-time picture
or cosmological language, i.e.\ we have to interpret the dynamical
constraint as an evolution equation \cite{PhD,cosmoIV}.

\section*{Acknowledgements}

The author thanks H.\ Kastrup for discussions and the
DFG-Graduierten-Kolleg ``Starke und elektroschwache Wechselwirkung bei
hohen Energien'' for a PhD fellowship.

\end{document}